\documentclass[12pt]{article}
\usepackage{graphicx}
\usepackage{epstopdf}
\usepackage{amsthm}
\usepackage{amsmath}
\usepackage{amssymb}
\usepackage{amstext}
\usepackage{amsfonts}
\usepackage{enumerate}
\usepackage[authoryear,nonamebreak,bibstyle]{natbib}
\usepackage{footnote}
\makesavenoteenv{tabular}
\makesavenoteenv{table}
\usepackage[onehalfspacing]{setspace}
\usepackage[shortlabels]{enumitem}
\usepackage{mathabx}
\usepackage{mathtools}

\usepackage{multirow}
\usepackage[title]{appendix}
\usepackage{xcolor}
\usepackage{cancel}
\usepackage{hyperref}
\usepackage[normalem]{ulem}
\usepackage{tikz}
\usetikzlibrary{shapes,arrows}
\usepackage{tikz-cd}
\usepackage{sectsty}
\usepackage{caption, subcaption}
\usepackage[numbib]{tocbibind}
\usepackage[colorinlistoftodos,textsize=tiny]{todonotes}

\begin{document}
\baselineskip=.22in\parindent=30pt

\newtheorem{tm}{Theorem}
\newtheorem{dfn}{Definition}
\newtheorem{lma}{Lemma}
\newtheorem{assu}{Assumption}
\newtheorem{prop}{Proposition}
\newtheorem{cro}{Corollary}
\newtheorem*{theorem*}{Theorem}
\newtheorem{example}{Example}
\newtheorem{observation}{Observation}
\newcommand{\exm}{\begin{example}}
\newcommand{\exmm}{\end{example}}
\newcommand{\obs}{\begin{observation}}
\newcommand{\obss}{\end{observation}}
\newcommand{\cor}{\begin{cro}}
\newcommand{\corr}{\end{cro}}
\newtheorem{exa}{Example}
\newcommand{\ex}{\begin{exa}}
\newcommand{\exx}{\end{exa}}
\newtheorem{remak}{Remark}
\newcommand{\rmk}{\begin{remak}}
\newcommand{\rmkk}{\end{remak}}
\newcommand{\thm}{\begin{tm}}
\newcommand{\nt}{\noindent}
\newcommand{\thmm}{\end{tm}}
\newcommand{\lm}{\begin{lma}}
\newcommand{\lmm}{\end{lma}}
\newcommand{\ass}{\begin{assu}}
\newcommand{\asss}{\end{assu}}
\newcommand{\df}{\begin{dfn}  }
\newcommand{\dff}{\end{dfn}}
\newcommand{\prp}{\begin{prop}}
\newcommand{\prpp}{\end{prop}}
\newcommand{\bqu}{\sloppy \small \begin{quote}}
\newcommand{\equ}{\end{quote} \sloppy \large}
\newcommand\cites[1]{\citeauthor{#1}'s\ (\citeyear{#1})}

\newcommand{\eq}{\begin{equation}}
\newcommand{\eqq}{\end{equation}}
\newtheorem{claim}{\it Claim}
\newcommand{\cl}{\begin{claim}}
\newcommand{\cll}{\end{claim}}
\newcommand{\bit}{\begin{itemize}}
\newcommand{\eit}{\end{itemize}}
\newcommand{\ben}{\begin{enumerate}}
\newcommand{\een}{\end{enumerate}}
\newcommand{\bcen}{\begin{center}}
\newcommand{\ecen}{\end{center}}
\newcommand{\fn}{\footnote}
\newcommand{\ds}{\begin{description}}
\newcommand{\dss}{\end{description}}
\newcommand{\prf}{\begin{proof}}
\newcommand{\prff}{\end{proof}}
\newcommand{\cs}{\begin{cases}}
\newcommand{\css}{\end{cases}}
\newcommand{\ml}{\item}
\newcommand{\lb}{\label}
\newcommand{\ra}{\rightarrow}
\newcommand{\tra}{\twoheadrightarrow}
\newcommand*{\supp}{\operatornamewithlimits{sup}\limits}
\newcommand*{\inff}{\operatornamewithlimits{inf}\limits}
\newcommand{\nf}{\normalfont}
\renewcommand{\Re}{\mathbb{R}}
\newcommand*{\mmax}{\operatornamewithlimits{max}\limits}
\newcommand*{\mmin}{\operatornamewithlimits{min}\limits}
\newcommand*{\argmax}{\operatornamewithlimits{arg max}\limits}
\newcommand*{\argmin}{\operatornamewithlimits{arg min}\limits}
\newcommand{\uhr}{\!\! \upharpoonright  \!\! }

\newcommand{\CR}{\mathcal R}
\newcommand{\CC}{\mathcal C}
\newcommand{\CT}{\mathcal T}
\newcommand{\CS}{\mathcal S}
\newcommand{\CM}{\mathcal M}
\newcommand{\CL}{\mathcal L}
\newcommand{\CP}{\mathcal P}
\newcommand{\CN}{\mathcal N}

\newtheorem{innercustomthm}{Theorem}
\newenvironment{customthm}[1]
  {\renewcommand\theinnercustomthm{#1}\innercustomthm}
  {\endinnercustomthm}
\newtheorem{einnercustomthm}{Extended Theorem}
\newenvironment{ecustomthm}[1]
  {\renewcommand\theeinnercustomthm{#1}\einnercustomthm}
  {\endeinnercustomthm}
  
  \newtheorem{innercustomcor}{Corollary}
\newenvironment{customcor}[1]
  {\renewcommand\theinnercustomcor{#1}\innercustomcor}
  {\endinnercustomcor}
\newtheorem{einnercustomcor}{Extended Theorem}
\newenvironment{ecustomcor}[1]
  {\renewcommand\theeinnercustomcor{#1}\einnercustomcor}
  {\endeinnercustomcor}
    \newtheorem{innercustomlm}{Lemma}
\newenvironment{customlm}[1]
  {\renewcommand\theinnercustomlm{#1}\innercustomlm}
  {\endinnercustomlm}
    \newtheorem{innercustomdf}{Definition}
\newenvironment{customdf}[1]
{\renewcommand\theinnercustomdf{#1}\innercustomdf}
{\endinnercustomdf}
\newtheorem{innercustomex}{Example}
\newenvironment{customex}[1]
{\renewcommand\theinnercustomex{#1}\innercustomex}
{\endinnercustomex}

\newtheorem{innercustomprp}{Proposition}
\newenvironment{customprp}[1]
{\renewcommand\theinnercustomprp{#1}\innercustomprp}
{\endinnercustomprp}

\newcommand{\red}{\textcolor{red}}
\newcommand{\blue}{\textcolor{blue}}
\newcommand{\purple}{\textcolor{purple}}
\newcommand{\mred}[1]{\color{red}{#1}\color{black}}
\newcommand{\mblue}[1]{\color{blue}{#1}\color{black}}
\newcommand{\mpurple}[1]{\color{purple}{#1}\color{black}}

\newcommand{\AG}[1]{\textcolor{magenta}{[AG:\ #1]}}
\newcommand{\MU}[1]{\textcolor{blue}{[MU:\ #1]}}

\makeatletter
\newcommand{\customlabel}[2]{%
\protected@write \@auxout {}{\string \newlabel {#1}{{#2}{}}}}
\makeatother


\def\qed{\hfill\vrule height4pt width4pt
depth0pt}
\def\reff #1\par{\noindent\hangindent =\parindent
\hangafter =1 #1\par}
\def\title #1{\begin{center}
{\Large {\bf #1}}
\end{center}}
\def\author #1{\begin{center} {\large #1}
\end{center}}
\def\date #1{\centerline {\large #1}}
\def\place #1{\begin{center}{\large #1}
\end{center}}

\def\date #1{\centerline {\large #1}}
\def\place #1{\begin{center}{\large #1}\end{center}}
\def\intr #1{\stackrel {\circ}{#1}}
\def\R{{\rm I\kern-1.7pt R}}
 \def\N{{\rm I}\hskip-.13em{\rm N}}
 \newcommand{\cprod}{\Pi_{i=1}^\ell}
\let\Large=\large
\let\large=\normalsize



\begin{titlepage}
\def\thefootnote{\fnsymbol{footnote}}
\vspace*{0.1in}

\title{Continuity Postulates and Solvability Axioms in \\ Economic Theory and in Mathematical Psychology:\\ \vspace{0.4em} A Consolidation of the   Theory of Individual Choice\fn{While acknowledging the pioneering works of Hugo Sonnenschein and Peter Wakker, the authors thank Yorgos Gerasimou, Alfio Giarlotta, Farhad Husseinov,   Rohit Parikh, Arthur Paul Pedersen, John Rehbeck, and Eddie Schlee   for stimulating conversations and correspondence. The authors also thank workshop participants at the \textit{Advances in Mathematical Economics} workshop at Universit\`a degli Studi di Napoli Federico II and seminar participants of the \textit{Seminar in Philosophy, Logic and Games} at CUNY. Finally, they thank an erudite referee of this journal for his/her suggestions, substantive and expositional; and also acknowledge discussions with Osama Khan way back in 2017. 
}}
\vskip 0.6em

\author   {Aniruddha Ghosh\fn{Department of Economics, Johns Hopkins University, Baltimore, MD 21218. {\bf E-mail} {aghosh23@jhu.edu}.}
 ~~ M. Ali Khan\fn{Department of Economics, Johns Hopkins University, Baltimore, MD 21218. {\bf E-mail} {akhan@jhu.edu}.}
 ~~  Metin Uyan{\i}k\fn{School of Economics, University of Queensland, Brisbane, QLD 4072.  {\bf E-mail} {m.uyanik@uq.edu.au}.}}



\date{\today}

\vskip 2.50em
%

\baselineskip=.18in

\noindent{\bf Abstract:}  This paper presents four theorems that connect continuity postulates in mathematical economics to solvability axioms in mathematical psychology, and ranks them under alternative supplementary assumptions.  Theorem \ref{thm_basic}  connects notions of continuity (full, separate, Wold, weak Wold, Archimedean, mixture) with those of solvability (restricted, unrestricted)  under the  completeness and transitivity of a binary relation.   Theorem \ref{thm_bequiv} uses the primitive notion of a separately-continuous function to answer the question when an analogous property on a relation is fully continuous. Theorem \ref{thm_fequiv} provides  a portmanteau theorem on the equivalence between restricted solvability and various notions of continuity under weak monotonicity. Finally, Theorem 4 presents a variant of Theorem 3 that follows Theorem 1 in dispensing with the dimensionality requirement and  in providing partial equivalences between solvability and continuity notions. These theorems are motivated for their potential use in representation theorems.  \hfill(137 words)



%

\vskip 0.50in


\noindent {\it Journal of Economic Literature} Classification
Numbers: C00, D00, D01

\medskip

\noindent {\it 2020 Mathematics Subject} Classification Numbers: 26B05, 91B02, 91B08, 91B42, 06A06

\medskip

\noindent {\it Key Words:}  Wold-continuity, separate continuity, restricted solvability, conjoint measurement.

\bigskip


\end{titlepage}

\large

\tableofcontents

\bigskip

\bcen  --------------  \ecen

\bigskip

\setcounter{footnote}{0}

\newpage

\large

\bqu The primary empirical meaning of continuity with respect to a connected product topology is precisely the solvability condition that it implies. It is more natural to state this empirical meaning explicitly, than to reinforce it into a stronger condition, of which the further empirical implications for finite data sets are not clearly identified.\\
\null\hfill{\cite[p.410]{kw03}}\equ

\bqu A number of results hold for much weakened versions of solvability, but many of the most important ones assume it and are not valid without it. For a time we believed that weakening the solvability assumption was largely a technical matter and that the results would differ little from the unrestrictedly solvable case. We now know this to be false, and in some of our future research we hope to gain a much deeper understanding of the possibilities that arise with weaker forms of solvability.  \null\hfill{\cite[p.47]{ln83}}\equ

\section{Introduction}\lb{sec: introduction} 


This paper is motivated by two themes. First, by integrating the various fragmentary usages  of the solvability and continuity axioms across the two disciplines of mathematical psychology and mathematical economics, respectively, we aspire to create a \textit{unitary}  discourse of  properties that have  taken  different forms across the two disciplines. We weave these fragments into a coherent set of results and aspire to bring together two communities which should not have moved away to begin with. Second, the analysis presented in this paper contribute to Luce-Narens' conceived research agenda of a systematic investigation of the \textit{possibilities} with weaker forms of solvability (for eg., restricted solvability) in 1983. While much work has been done on forms of solvability and its implications for additive utility representation,\fn{See \cite{go97,go00,go03} for details.} we believe that a deeper investigation of the \textit{form} and \textit{content} of the solvability axioms has been unexplored and  its intimate connections with the continuity postulates in mathematical economics most certainly remain uninvestigated to the extent that they ought. The \textit{leitmotif} of this paper, therefore, is to go back to the most raw, primitive and undefined, meanings of the terms solvability and continuity,  and  establish analytical and substantive connections between them and their  subsequent conceptual proliferations.\fn{We refer the reader to \cite{kw03} for a comprehensive introduction to the axioms of continuity and solvability across the topological and the algebraic registers.}

The axiom of solvability has been a cornerstone assumption of the measurement theory literature since it was first introduced in the seminal work of \cite{lt64}. The idea of the existence of solutions to equations for ``fundamental quantities"\fn{See \citet[Chapter 1]{klst71} 
 for details as to what the subject regards as \lq\lq fundamental quantities''.} dates back at least to Helmholtz (1887), but it wasn't until \cite{ho01} who formally introduced the axiomatic approach to measurement theory.\fn{See \cite{me96} for the translated version of Part 1 of H\"older's text. Also see \cite[p. 300]{bi67},  \cite[p. 45]{fu63},  \cite[p. 54]{klst71} and the comprehensive text  \cite{mo16}; especially Chapters 8 and 15 on Stevens and Suppes respectively therein.} With Luce and Tukey's 1964 axiomatization and its culmination in the 1971 treatise \textit{Foundations of Measurement} (\citealp{klst71}), the solvability axiom was  concretized  in mathematical psychology. The motive was to impose enough \textit{richness} on the \textit{algebraic} structure of the space concerned such that solutions to certain equations can be found. Solvability was then categorized in two forms, a restricted and an unrestricted version,  with most of the applications in the literature working with the restricted version. Much of our work in this paper investigates the restricted solvability axiom while establishing the strength of the  unrestricted solvability axiom.





 

 


Just as with the solvability axioms in mathematical psychology,  continuity postulates in mathematical economics are used to enforce enough richness on the \textit{topological} structure being investigated.  However, in the absence of a coherent and unified formulation between the two conceptions, different versions of both the notions are invoked in different settings.\fn{In this paper, we do not investigate the behavioral implications of continuity and solvability. See \cite{ku21et} for  a two-way relationship between connectedness and behavioral assumptions on preferences.}  The literature is replete with the usage of the adjective \textit{continuous}, applied  to a function as well as to a binary relation. In keeping with the theme of understanding the \textit{richness} of a structure, we  detail  the intricacies of the different conceptions of continuity of a binary relation as comprehensively investigated in \cite{uk19a} with a special emphasis on {\it full, separate, Wold,  weak Wold, Archimedean and mixture-continuous} manifestations. This work constitutes an important backdrop for carrying out the investigation pursued  in this paper.





With this framing of the project and its underlying motivation in place, we can now turn to the results themselves and inquire how they contribute to mathematical psychology and mathematical economics. Theorems 1, 2, 3 and 4 in Section 4 are the main results, with Theorems 1 and 4 dispensing with the dimensionality requirement of the choice space.  Theorem \ref{thm_basic} presents some well-known and some new results  by connecting continuity with solvability, the latter being of primary interest for mathematical psychologists working in abstract decision theory. Along with five examples, it exhaustively documents connections between and across various notions of continuity  and solvability  under the maximally parsimonious  assumptions on a binary relation. Theorem \ref{thm_bequiv} takes the primitive notion of a separately-continuous function and uses it to define separate continuity of a relation and delineates conditions under which it is fully continuous in a finite-dimensional setting. Theorem \ref{thm_fequiv} provides, under weak monotonicity,  a portmanteau theorem on the equivalence between the six notions of continuity and restricted solvability of a binary relation with Theorem 4 providing partial equivalences for the infinite dimensional setting. It is our hope that the theorems presented will further the aims of Luce and Narens' research agenda, as elucidated in the epigraph. Similarly, we aim to complement \cite{kw03} by providing a ranking of the continuity and solvability axioms which we hope serves as a go-to reference for future representation theorems.\fn{\cite{wy19jet, wy21ime} provide powerful tools for analyzing concavity and convexity of utility and weighting functions by using the convexity of preferences without hinging on the continuity assumption.}




The rest of the paper is organized as follows. Section 2 recapitulates some well-known solvability axioms  and continuity postulate in mathematical psychology  and  mathematical economics, respectively. Section 3   exhaustively links the two axioms and presents our main results. In Section 4, we present six applications of our work spanning, (i) Walrasian equilibrium theory, (ii) consumer theory, (iii) representation of preferences in economics, and  (iv)  mathematical psychology. Finally, Section 5 concludes with some observations on future research directions. Appendix A provides the proofs of the results, while Appendix B pins down some observations that arise in the course of the results.

\vskip0.2cm

\section{The Background: A Recapitulation}

\subsection{The Continuity Postulate in Mathematical Economics}


In this section, we describe what it means for a binary relation to be Archimedean, continuous, mixture-continuous, separately continuous, weak Wold-continuous and Wold-continuous, based on a recapitulation of \cite{uk19a}.


Let $X$ be a set. A {\it (binary) relation} $\succsim$ on $X$ is a subset of $X\times X$. The {\it  asymmetric part} $\succ$ of $\succsim$ is defined by $x\succ y$ if $x\succsim y$ and  $y\not\succsim x$, and its {\it symmetric part} $\sim$ is defined by $x\sim y$ if $x\succsim y$ and $y\succsim x.$   We call $x\bowtie y$ if $x\not\succsim y$ and $y\not\succsim x$. For any $x\in X$, let   $A_\succsim(x)=\{y\in X| y\succsim x\}$ denote the {\it upper section} of $\succsim$ at $x$  and  $A_\precsim(x)=\{y\in X| y\precsim x\}$ its {\it lower section} at $x$.    %
%
 
 %

We next provide topological properties of $\succsim$ provided that $X$ is endowed with a topology; for topological concepts that are undefined in this paper, see \cite{wi70}. The relation $\succsim$ has {\it closed (open) graph} if it is closed (open) in the product space $X\times X$ endowed with the product topology; has {\it closed (open) upper sections} if its  upper sections are closed; has {\it closed (open) lower sections} if its  lower sections are closed;  and has {closed (open) sections} if it has closed (open) upper and lower sections. Moreover, $\succsim$  is {\it continuous} if it has closed sections and its asymmetric part $\succ$ has open sections.    

 The following two definitions are not standard in the literature and hence, we provide them separately.

\df{\nf(Restricted Continuity)}
Let $\succsim$ be a binary relation on $X$ and $S\subseteq X$. The relation $\succsim$ on $X$ is {\it restrictedly continuous with respect to $S$} if for all $x\in X$, $A_\succsim(x)\cap S \text{ and } A_\precsim(x)\cap S \text{ are closed in } S, \text{ and } A_\succ(x)\cap S \text{ and } A_\prec(x)\cap S \text{ are open in } S$. 
\label{df:restcont}
\dff
  
\noindent We sometimes use the following compact version for the restricted continuity of a binary relation $\succsim$ given a set $S$: {\it the restriction of $\succsim$ on $S$ is continuous}, or simply {\it $\succsim \upharpoonright \!\! S $ is continuous.} 
The restricted continuity concept is not an \textit{ad hoc} one: it is not difficult to show that for a complete binary relation, it is a generalization of the mixture continuity concept, defined below as in \cite{hm53}; see \cite{uk19a} for a derivation and discussion, and also \citet[Section 3]{re08bulletin} for different versions of  restrictions of binary relations that are used in mathematics. 

\df{\nf(Separate-Continuity)}
 Let $I$ be an arbitrary indexed set, $X_i$ a non-empty subset of $\Re$ for all $i\in I$,  $\hat X= \prod_{i\in I}X_i$ endowed with the product topology and $X\subseteq \hat X$  a convex set.   A relation $\succsim$ on $X$   is  {\it separately continuous} if for all $i\in I$ and $x\in X$, the restriction of the relation on $L_{i,x}=\{(y_i, x_{-i})\in X|y_i\in X_i\}$ is continuous, where  $(y_{i},x_{-i})$ denotes $y_{i}\in X_{i},x_{-i}\in X_{-i}.$ . \label{df:separate}
\dff

\nt This definition is motivated by the well-known separate continuity of functions (requiring continuity in each coordinate separately) that goes back to Cauchy's classic work; we return to this in the next section.   
\smallskip

  Now let $X$ be a convex subset of a linear space and for all $\lambda \in [0,1]$ and all $x,y\in X$, $x\lambda y$ denotes $\lambda x+ (1-\lambda)y$. A relation $\succsim$  on $X$ is  
{\it  mixture-continuous} if $x,y,z\in X$ implies that the sets $\{\lambda\in [0,1]| x\lambda y\succsim z\}$ and  $\left(\{\lambda\in [0,1]| x\lambda y\precsim z\} \right)$ are closed, and   $\succsim$ is    {\it  Archimedean} if $x,y,z\in X, x\succ y$ implies that there exist $\lambda, \delta\in (0,1)$   such that $x\lambda z\succ y$ and $x\succ y\delta z$.  
The above two postulates are defined by using straight lines, whereby a \textit{straight line} in $X$ is the intersection of a one dimensional affine subspace of $\Re^I$ and $X$. Next, we generalize the notion of straight lines to unbroken curves following the seminal work of \cite{wo43}.  
Now, let $X$ be a topological space. An {\it arc} in   $X$ is a continuous injective function $m:[0,1]\ra X$.  
A {\it curve} in $X$ is the image of an arc $m: [0,1]\ra X$.  
Since an arc $m$ is continuous and injective, it is a bijection from $[0,1]$ to its image $m([0,1])$. When $X$ is Hausdorff, it follows from $[0,1]$ being compact and $m([0,1])$ being Hausdorff that $m$ is a homeomorphism between $[0,1]$ and $m([0,1]).$ Hence, these two spaces are homeomorphic; see for example  
\citet[Theorem 17.14, p. 123]{wi70}. Note  that an arc induces a unique curve but a curve can be induced by distinct arcs: for example, any closed segment $[x,y]$ of the diagonal in $\Re^2$ is induced by every arc with $m_1(\lambda)=m_2(\lambda)$ for all $\lambda\in [0,1]$ where $m_1(0)=x, m_1(1)=y$.

%

Next, we define what it means for a relation to be order-dense and present the concepts of Wold and weak Wold-continuity. 

\df{\nf(Order Dense)}
A relation $\succsim$ on a set $X$ is {\it order dense} if $x\succ y$ implies that there exists $z\in X$ such that $x\succ z\succ y$.
\dff

\nt The definition of an order-dense relation implies a certain \textit{richness} of the space such that an element in the space can always be found that is `sandwiched' (by order) between two given elements.



\df{\nf(Wold-solvability/Wold-continuity)}
Let $X$ be a convex subset of a topological vector space. A binary relation $\succsim$ on $X$ is 
 \ben[{\nf (i)}, topsep=1pt]
 \setlength{\itemsep}{-2pt} 
 \ml  {\it Wold-solvable} if $x\succ z\succ y$ implies that any curve joining $x$ to $y$ meets the indifference class of $z$.
 \ml {\it weak Wold-solvable} if $x\succ z\succ y$ implies that the straight line joining $x$ to $y$ meets the indifference class of $z$. 
 \ml {\it Wold-continuous} if it is order dense and Wold solvable. \label{it_wc} 
 \ml  {\it weakly Wold-continuous} if it is order dense and weak Wold-solvable.\label{it_wwc} 
 \een
\label{df: wold}
\dff

\rmk{\nf 
Part \ref{it_wc}  is Axiom 5 in \cite{wo43} and part \ref{it_wwc} is Axiom B in \citet[p. 82]{wj53}. For the latter, note that order denseness follows from  their monotonicity assumption (Axiom A), hence under their monotonicity assumption, weak Wold-solvablility and weak Wold-continuity are equivalent. 
}\rmkk

  We next provide examples of binary relations that demonstrate the above definitions. 
\ex{\nf(\cite{gp84})}
\nf Let the binary relation $\succsim$ be defined on $\Re^{2}$ by $(x_{1},x_{2})\succsim (y_{1},y_{2}) \text{ iff } f(x_{1},x_{2})\geq f(y_{1},y_{2})$ where 
\[
f(x_{1},x_{2}) = 
\begin{cases}
\dfrac{x_{1}x_{2}}{x_{1}^{2}+x_{2}^{2}} &  (x_{1},x_{2})\neq (0,0), \\
\ 0 & (x_{1},x_{2}) = (0,0).  \\
\end{cases}
\]

\nt This is the textbook example of a separately continuous function which is not continuous; it was originally presented by  \cite{gp84}. The relation is  continuous along any restriction parallel to either of the axes and hence, separately continuous. However, it is discontinuous along the $45^{\circ}$  line (at $(0,0)$) and therefore, not continuous. Separate continuity of a binary relation is indeed  weaker than continuity. \qed 

\exx

\ex 
\nf As in Example 1,  define  $\succsim$ on $\Re^{n}$ as  $(x_{1},x_{2},...,x_{n})\succsim (y_{1},y_{2},...,y_{n}) \text{ iff } x_{1}+x_{2}+...+x_{n}\geq y_{1}+y_{2}+...+y_{n}$. 
The relation  $\succsim$  is order dense because between any $x$ and $y$ there exists $z$ such that $x\succ z \succ y.$  
Moreover, this relation is Wold-continuous as any continuous arc joining two points in $\Re^{n}$ intersects indifference curve of a point sandwiched between the two as per the binary relation.
\qed
\exx

\ex 
\lb{lexicographic}
\nf Let $X= \Re^{2}.$ Define $\succsim$ on $X$ as 
 $(x_{1},x_{2})\succsim (y_{1},y_{2})\text{ iff } x_{1}\geq y_{1}\ or \ x_{1}=y_{1}, x_{2}\geq y_{2}$.  
 The relation $\succsim$  is called {\it lexicographic}. It is trivial to see that $\succsim$ is order-dense. In order to see that $\succsim$ does not satisfy weak Wold-continuity, note that  $(\frac{1}{2},1) \succsim (\frac{1}{3},0) \succsim (0,1)$. For $t\in [0,1]$, we have, $t(\frac{1}{2},1)+(1-t)(0,1)=(\frac{1}{2}t,1)$. But for no $t\in [0,1],$ we have  $(\frac{1}{2}t,1)\sim (\frac{1}{3},0)$. Therefore, it is not weak-Wold solvable and hence, not weak Wold-continuous. Therefore, it is also not Wold-continuous. \qed
\exx

\ex
\nf Let the binary relation $\succsim$ be defined on a bounded subset $X\subset\mathbb R^{2}$ as $(x_{1},x_{2})\succsim (y_{1},y_{2})$ if and only if $f(x_{1},x_{2})\geq f(y_{1},y_{2})$  
where 
\[
f(x_{1},x_{2}) = 
\begin{cases}
0.4, & \text{ if } x_{1}+x_{2}\leq 1, (x_{1},x_{2})\neq (1,0), \\
0.5, & \text{ if } (x_{1},x_{2})= (1,0),\\
0.6, & \text{ if }x_{1}+x_{2}> 1. \\
\end{cases}
\]
It can be easily verified that a relation defined this way is neither order-dense nor weak-Wold solvable (Wold solvable).

\exx

We now present a theorem due to  \citet[Theorem 2]{uk19a} that establishes equivalence between various notions of continuity of a binary relation on finite dimensional Euclidean spaces. But first we describe some preliminary properties of the binary relation. 
A relation  $\succsim$ is {\it strongly monotonic} if   $x> y$ implies $x\succ y$ and {\it weakly monotonic} if for all $x,y\in X$, $x>y$ implies  $x\succsim y$, and  $\succsim$ is convex if it has convex upper sections. For vectors $x$ and $y$,  we use the following convention: ``$x\geq y$" means $x_i\geq y_i$  in every component,  ``$x> y$" means $x\geq y$ and  $x\neq y$, and ``$x\gg y$" means $x_i>y_i$  in every component. We call a subset $A$ of $X$ is {\it bounded by $\succsim$}   if for all $x,y\in A$ there exists $a,b\in X$ such that $a\succsim x,y$ and $x,y\succsim b$. A {\it polyhedron} is a subset of $\Re^n$ which is an intersection of a finite number of closed half-spaces.

\medskip

\begin{customthm}{\nf(Uyanik-Khan -- Finite Dimension)}
{\it Let $\succsim$ be a complete and transitive relation on a convex subset $X$ of $\Re^n$. If either {\nf (i)} $\succsim$ is convex and $X$ is either a polyhedron or open, or {\nf (ii)} $\succsim$ is weakly monotonic and $X$ is  bounded by the usual relation $\geq$, then the following   continuity postulates for $\succsim$ are equivalent: graph continuous, 
continuous,    
mixture-continuous, 
Archimedean,   
Wold-continuous, and    
weakly Wold-continuous. }

\end{customthm}

\medskip

\nt Under the convexity postulate, the equivalence result fails for infinite dimensional spaces; see the example in \citet{nt95}. However, Khan-Uyanik (2019) show that some of the equivalences pertaining to scalar continuity postulates in the Theorem above do not require finite-dimensionality or a topological structure on the choice set and prove   
\medskip

\begin{customthm}{\nf(Uyanik-Khan -- Infinite Dimension)}
{\it 
For every convex, complete and transitive relation on a convex subset of a vector space, the following continuity postulates are equivalent: Archimedean,  
strict Archimedean, mixture-continuous and weakly Wold-continuous. 
}
\end{customthm}
\medskip

\nt We show below that under the monotonicity postulate, the portmanteau equivalence theorem holds for infinite dimensional spaces and also extends to the separate continuity and the restricted solvability postulates.

\subsection{The Solvability Axiom in Mathematical Psychology}

In this subsection, we consider two solvability notions as articulated in the measurement theory literature by \cite{klst71}. They categorize solvability as a structural axiom asserting that ``solutions exist to certain classes of equations or inequalities;" see \cite{ln86} for a review of generalizations of classical theories of measurement. An axiom is said to be \textit{structural} if it restricts the set of structures satisfying a given axiom system to something less than the set determined by the representation theorem; i.e., a tighter bound is imposed by structural axioms on admissible structures. We refer the reader to  \cite{ps74} for a discussion on structural axioms in the context of theories of subjective probability.
%

Throughout this subsection, we assume  $I$ is an arbitrary indexed set, $X_i$ a non-empty set for all $i\in I$,  $\hat X= \prod_{i\in I}X_i$ and $X\subseteq \hat X$.  

\df{\nf(Unrestricted Solvability)}
A relation $\succsim$ on $X$ satisfies unrestricted solvability with respect to the $i^{th}$ component if for any $x\in X$ and $y_{-i}\in X_{-i}$, there exists $z_{i}\in X_{i}$ such that $x\sim (z_{i}, y_{-i} ).$ When this holds for all $i\in I$, the binary relation is said to satisfy unrestricted solvability.
\label{df_us_inf}
\dff



The following example illustrates a binary relation that is unrestricted solvable on $\Re^{n}$. 

\begin{customex}{1 (\nf contd)}
\nf Let $ X= \Re^{n}, n\geq 2.$ Define $\succsim$ on $X$ as $(x_{1},x_{2},...,x_{n})\succsim (y_{1},y_{2},...,y_{n})$ if and only if  $x_{1}+x_{2}+...+x_{n}\geq y_{1}+y_{2}+...+y_{n}$. 
Hence, $(x_{1},x_{2},...,x_{n})\sim (y_{1},y_{2},...,y_{n})$ iff  $x_{1}+x_{2}+...+x_{n}=y_{1}+y_{2}+...+y_{n}.$ For any given $2n-1$ out of the $2n$ variables, there exists a value for the $2n$-\nf th variable such that the indifference of the binary relation is met. 
\qed
\end{customex}

\df{\nf(Restricted Solvability)} A relation $\succsim$ on $X$ satisfies restricted solvability with respect to the $i^{th}$ component if for any $x\in X$, $a_{i}, b_{i}\in X_{i}$, $y_{-i}\in X_{-i}$  with  $(a_{i},  y_{-i} )\succsim x\succsim (b_{i}, y_{-i} )$, there exists $c_{i}\in X_{i}$ such that $x\sim (c_{i}, y_{-i} ).$ When this holds for all $i\in I$, the binary relation is said to satisfy restricted solvability. 
\label{df_rs_inf}
\dff


In the following example, restricted solvability fails to hold. The failure to meet restricted solvability comes from one of the components not being `dense' enough.

\ex
{\nf  Let $ X_{1}=\Re$ and $ X_{2}=\mathbb Q.$ 
 Define $\succsim$ on $\Re \times \mathbb Q$ as 
 $(x_{1},x_{2})\succsim (y_{1},y_{2})\quad\text { iff }\ x_{1}+x_{2}\geq y_{1}+y_{2}$.   
Hence, $(0,2)\succsim (\sqrt{2},0) \succsim (0,0) $ but there is no $x_{2}\in \mathbb Q$ such that $(0,x_{2})\sim(\sqrt{2},0)$. Therefore, restricted solvability fails for the second component. However, restricted solvability holds for component 1.}
\qed

\exx

 \section{Main Results}

 In this section, we present four theorems, the first and the third substantial, and the second a mathematical result that links the two by bringing in insights from the mathematics literature. Theorem \ref{thm_basic} documents connections between and across various notions of continuity (full, separate, Wold, weak-Wold, mixture-continuity and Archimedean) and solvability (restricted and unrestricted), and Theorem \ref{thm_fequiv} provides a portmanteau equivalence theorem under the weak monotonicity assumption. Theorem \ref{thm_bequiv} borrows the idea of separate continuity of functions from the  literature on continuity of functions and provides conditions under which separate continuous relation is fully continuous for finite dimensions. Theorem 4 provides a partial equivalence theorem for infinite dimensional spaces.


 

 
 
 
 

  \thm
\label{thm_basic} 
 Let $I$ be an arbitrary indexed set, $X_i$ a non-empty subset of $\Re$ for all $i\in I$,  $\hat X= \prod_{i\in I}X_i$ endowed with the product topology and $X\subseteq \hat X$  a convex set.     Then the following implications hold for a complete and transitive binary relation $\succsim$ defined on $X$.
 \ben[{\nf (a)}, topsep=1pt]
 \setlength{\itemsep}{-2pt} 
 \ml  Continuity  $\Rightarrow$ Wold-continuity $\Rightarrow$ weak Wold-continuity $\Rightarrow$   restricted solvability and Archimedean. 
 \ml Continuity  $\Rightarrow$ mixture-continuity $\Rightarrow$ separate continuity,  weak Wold-continuity  and Archimedean. 
 
  \ml Separate continuity $\Rightarrow$   restricted solvability.    
 \ml Unrestricted solvability $\Rightarrow$ restricted solvability.
 \een
    
 \thmm

\noindent  Theorem \ref{thm_basic} raises as many questions as it answers. In particular  there are no  further relationships among these six continuity postulates under the least restrictive assumptions of completeness and transitivity. 
  We use examples in  Appendix \ref{sec_examples} to establish this claim. 
 
Next, we turn to our second result.

\thm
Let $\succsim$ be a complete and transitive  preference relation that is weakly monotone in $n-1$ of its coordinates on a convex, order-bounded set $X$ in $\Re^{n}$. Then, on  {\nf int}$X$, $\succsim$ is  separately continuous if and only if it is continuous, where {\nf int}$X$ is the interior of the set $X.$  
\label{thm_bequiv} 
\thmm




\nt  
This result provides an alternative proof to the existing results for functions since a preference relation which is continuous, complete and transitive on subsets of $\Re^n$ is representable by a continuous function, a version of this result for functions is provided by \cite{yo10qjpam} for $n=2$ and generalized to any finite $n$ by \cite{kd69amm}; see \cite{cm16} for a detailed discussion. However, note that since it is not known if a separately continuous binary relation is representable by a separately continuous function, this result does not follow from the existing results on functions.\fn{We thank an anonymous referee for emphasising this point.}  Moreover, the completeness assumption is redundant in this theorem. Towards this end, we present the proof without the completeness assumption. In fact, if the choice set $X$ has at least two strictly comparable points, then the relation is complete; see \citet[Theorem 2]{ku21et}, \cite{gw20} and \cite{ku20a} for further details. Note that the interiority assumption is not redundant in the theorem above; Example 6 in the appendix shows that this theorem is not true without the interiority assumption.

 
 
 


We now present an equivalence theorem for finite dimensional spaces. 

  \thm
  \label{thm_fequiv}
 Let $\succsim$ be a complete, transitive, order dense and weakly monotonic binary relation on a convex and order-bounded set $X$  in $\mathbb R^n.$ Then, on {\nf int}$X$, the following postulates are equivalent for $\succsim$: continuity, Wold-continuity, weak Wold-continuity, mixture-continuity, Archimedean, separate continuity, restricted solvability.  
    \thmm

Figure 1 illustrates the relationship among different continuity postulates. The black arrows illustrate  the relationships presented in Theorem 1, the red and green arrows illustrate the additional relationships under weak monotonicity for finite dimensional spaces presented in Theorems 2 and 3, and the green arrows illustrates additional relationships under weak monotonicity for possibly infinite dimensional spaces presented in Theorem 4.  

\begin{figure}[!ht]
	\begin{center}
		\includegraphics[width=6.7in, height=2.05in]{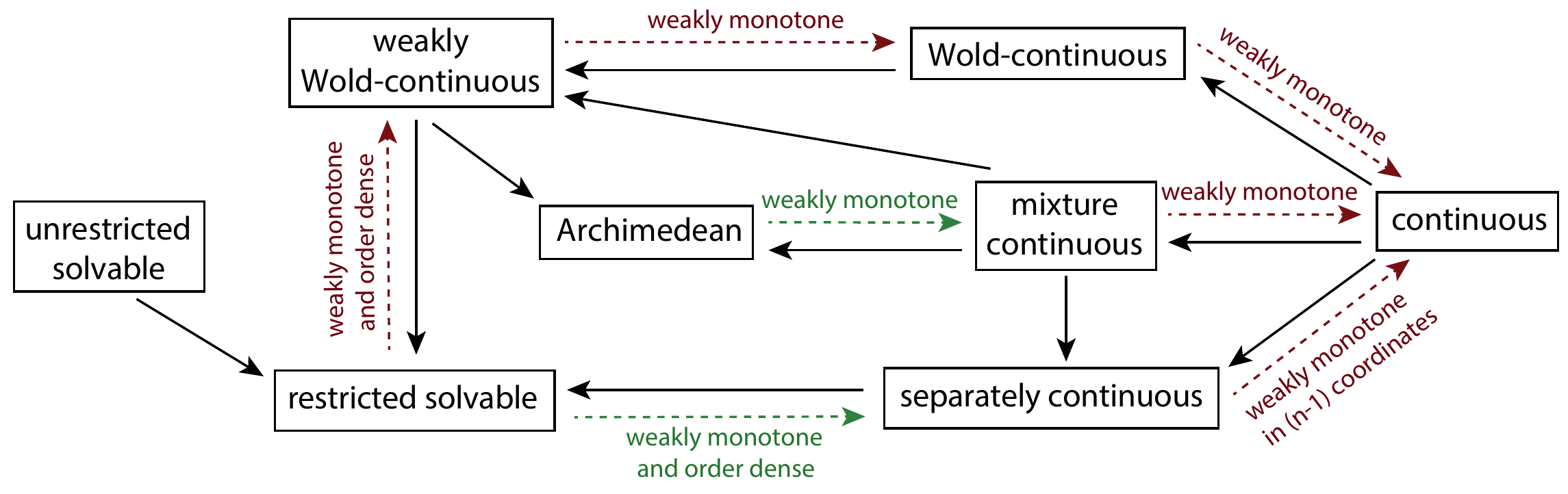}
	\end{center}  
	\vspace{-15pt}   
	
	\caption{Solvability and Continuity}
	\vspace{-5pt}   
	
	\lb{fig: relation}
\end{figure}

\noindent Unrestricted solvability is a strong conception of continuity. This is seen in Figure 1 where there are no incoming arrows from the rest of the continuity postulates to the unrestricted solvability postulate even under the assumptions of convexity  and weak monotonicity; see Example 1 in Appendix \ref{sec_examples} for details. In terms of \textit{strength} of these axioms, under the set of assumptions employed in this paper, restricted solvability is the \textit{weakest} and continuity is the \textit{strongest} and everything else is in between. 
 \smallskip

We 
now highlight  the importance of the weak monotonicity assumption in Theorem \ref{thm_fequiv}.    \citet[Theorem 2]{uk19a} provide an equivalence theorem for convex or weakly monotone preferences. In their theorem, solvability and separate continuity postulates are missing. Theorem \ref{thm_fequiv} includes these two continuity postulates in the picture and shows that for weakly monotonic preferences, an extended equivalence can be established.  However, for convex preferences, without the monotonicity assumption, restricted solvability does not imply other continuity assumptions; see Example 2 in  Appendix \ref{sec_examples}.

Theorems 2 and 3 are true for finite dimensional spaces. We next present an example showing that the equivalence relationships need not be true for infinite dimensional spaces.

\ex 
\label{exm_infinite} 
 {\nf Let $X_i=(-10, 10)$, $I=\mathbb Z_+$, $X=\prod_{i\in I} X_i$ endowed with the product topology and $u: X\ra \Re$ is defined by $u(x)=\inf_{i\in I}\{x_i\}$.  
 Let $\succsim$ be the  preference relation induced by $u$; that is, $x\succsim y$ if and only if $u(x)\geq u(y)$. 
 
 It is clear that the restriction of  $u$ on any line parallel to a coordinate axis is continuous since for any $i$ and $x_{-i}$, $u(x_{-i}, y_{i})=y_i$ for all $y_i\in X_i$. Moreover, $u$ is monotone by construction. Then, it is easy to see that $\succsim$ is monotone and separately continuous.    Finally, $\succsim$ is not continuous. In order to see this, let $x=(1, 1, \ldots)$, $y=(2,2,\ldots)$ and $y^n$ be defined by follows: $y^0=(0,0,\ldots)$, $y^1=(2, 0, 0,\ldots)$, $y^2=(2,2,0, \ldots)$, \dots. It is not difficult to show that $y^n$ converges $y$ (in the product topology); see for example Munkres (2000, Exercise 6, p. 118) for characterization of the convergence of a sequence in a product space.  It is clear that $x\succsim y^n$ for all $n$ and $y\succ x$. Hence, $\succsim$ is not continuous.    
Next, we show that $\succsim$ is also mixture continuous.  
Towards this end, assume there exists $x,y,z\in X$ and $\lambda^n \ra \lambda\in [0,1]$ such that $x\lambda^n y\succsim z$ for all $n$ but $x\lambda y \prec z$. Hence, inf$_i (\lambda^n x_i +(1-\lambda^n) y_i) \geq \text{\nf inf}_i z_i$ and inf$_i (\lambda x_i +(1-\lambda) y_i) < \text{\nf inf}_i z_i$. Then, for all $\lambda'$ in some small neighborhood of $\lambda$, inf$_i ( \lambda' x_i +(1-\lambda') y_i) < \text{\nf inf}_i z_i$.   
 This furnishes us a contradiction.     
} \qed 
\exx

We  end this section with the following partial equivalence result for infinite dimensional spaces.


   
  \thm
  \label{thm_fequivInf}
 Let $\succsim$ be a complete, transitive, order dense and weakly monotonic binary relation on a convex  set $X$ $\subseteq$ $\prod_{i\in I}X_i, X_{i}\subset \mathbb R, \forall i\in I$ such that for all $x\in X$ there exist  $\varepsilon>0$ such that $(x_i -\varepsilon)_{i\in I}, (x_i +\varepsilon)_{i\in I}\in X$ (strong order-boundedness). Then the following postulates are equivalent for $\succsim$: weak Wold-continuity, mixture-continuity, Archimedean. Moreover,  the following two are equivalent: separate continuity and  restricted solvability. 
    \thmm

 \rmk\label{rmk_infexamples}
{\nf This theorem provides a partial relationship among different continuity postulates in infinite dimensional spaces. Note that the preference relation in Example 5 for an infinite dimensional space is separately continuous and mixture continuous but not fully continuous, and hence shows that neither separate continuity, nor mixture-continuity  implies full continuity. 
 We refer the reader to Figure 1 to work out for herself the additional supplemental relationships that need to be worked out. And we leave the investigation of these additional relationships   for infinite dimensional spaces as an open problem.} 
\rmkk

\rmk {\nf  
The results may be generalized to spaces $X$ where $X_i$ is a convex subset of a completely ordered topological vector space for all $i\in I$. But the interaction between order and topological vector space structure needs careful attention.  A useful result in this context is one by \cite{fl61}: {\it A linearly ordered set $X$ that is topologically separable in its order topology and has countably many jumps is order-isomorphic to a subset of the real numbers}. That is,  when the components of the Cartesian product is a linearly ordered separable space and the binary relation is continuous, then it is isomorphic to an interval in the real line, hence the generalization of our results to such a setting could potentially be done; see also \cite{me86}. Another question here is whether weaker concepts such as separate continuity/solvability are enough to obtain isomorphism of an interval. We leave these to future investigation.} 
\rmkk

\section{Some Selective Applications}


In this section, we are very much motivated by the stance of \cite{kw03} on the empirical content imbibed in the continuity and solvability axioms. This motivates us to re-work the representation theorems since the move from solvability to continuity is an essential cog of our results; something the literature hasn't been familiar with yet. We enlist applications that span mathematical psychology, consumer theory and general equilibrium. 
 
\subsection{Wold's (1943) pioneering representation of preferences}

\cite{wo43} proves the following utility representation theorem. 

\smallskip 

\prp{\nf(Wold Representation)} {\it  
 Every complete, transitive, weakly monotonic and Wold-continuous preference relation on $\Re_+^n$ is representable by a weakly monotonic and continuous utility function.} 
\smallskip
 
  Theorem \ref{thm_fequiv} shows that it is possible to replace the continuity assumption of Wold by any of the continuity assumptions listed in Theorem \ref{thm_fequiv}. \cite{wj53} replace weak monotonicity by strong monotonicity and Wold-continuity with weak Wold-continuity. Hence, Theorem \ref{thm_fequiv} applies to this result of Wold-Jureen. 

\prpp

\subsection{Aumann's (1966) existence theorem on Walrasian equilibria}  

\cite{au66} proves a theorem on the existence of a Walrasian equilibrium in an exchange economy with continuum of consumers. Aumann's exchange economy is defined by follows. 

\df
An exchange economy is a list $\mathcal E=(T, \{X_t\}_{t\in T}, \{\succsim_t\}_{t\in T},  e)$ where 
\ben[{\nf (a)}, topsep=3pt]
\setlength{\itemsep}{-1pt} 
\ml $T=[0,1]$ is the set of consumers with the lebesgue measure.  
\ml $X_t=\Re_+^n$ is the consumption set of consumer $t\in T$. 
\ml $\succsim_t$ is the preference relation of consumer $t\in T$ on her consumption set $\Re_+^n$.
\ml $e: T\ra \Re_+^n$ is a measurable function with $e_t$ denoting the endowment of consumer $t$. 
\een
\label{df_aumann}
\dff

An {\it assignment} in an economy is a measurable function $x: T\ra \Re_+^n$.  An {\it allocation} is an assignment $x$ such that $\int x=\int e$.  A {\it competitive equilibrium} is a  price-allocation pair $(p,x)\in \Re_+^{2n}$  such that for all consumer $t$, $x(t)$ is maximal with respect to $\succ_t$ in the ``budget set" $B_p(t)= \{x\in X_t: p \cdot x\leqq p \cdot e(t)\}$.


\medskip 
\prp{\nf(Aumann's Existence Theorem)}
\textit{Let $\mathcal E$  be an economy defined above such that 
\ben[{\nf (a)}, topsep=3pt]
\setlength{\itemsep}{-1pt} 
\ml $\int e >0$, and 
\ml for each $t\in T$, $\succsim_t$ is continuous, strongly monotonic and satisfies the following measurability assumption: if $x$ and $y$ are assignments, then the set $\{t: x(t) \succ_t y(t)\}$ is measurable. 
\een
Then the economy has a competitive equilibrium. }
\prpp
\smallskip

Under Aumann's strong monotonicity assumption, Theorem \ref{thm_fequiv} implies that it is possible to replace the continuity assumption of Aumann by any of the continuity assumptions listed in Theorem \ref{thm_fequiv}. 


\subsection{Wakker's (1989) investigation of additive separability } 

%

We next provide two applications from \cite{wa89}. Both applications replace continuity by restricted solvability under the most general conditions.

The following lemma presents conditions on the topological space under which a continuous binary relation implies restricted solvability, see Chapter 3, p.44 of Wakker (1989). 

\begin{customlm}{\nf(Wakker (1989))}
 \lb{wk1} 
 Under the topological assumption of connected separability such as $\Re^n$, a continuous weak order $\succsim$ implies restricted solvability.
\end{customlm}

Using Theorem \ref{thm_fequiv}, we can now state the lemma as an equivalence proposition.
\prp
Let $\succsim$ be a complete, transitive, order dense and weakly monotonic binary relation on a convex and order-bounded set $X$ in $\Re^{n}$. Then $\succsim$ is continuous if and only if it is restricted solvable.

\lb{wk2}
\prpp

Let $A \subseteq \{1,2,\ldots,n\}$ and let $x_{A}$ and $x_{\neg A}$ denote the restriction of $x\in \mathbb R^{n}_{++}$ to $A$ and $\neg A,$ respectively.\fn{If $n=3$ and $A=\{1,2\}$, then for $x=(x_{1},x_{2},x_{3})\in \mathbb R^{3}_{++},$ $x_{A}=(x_{1},x_{2})$ and $x_{\neg A}=(x_{3})$ are  the restriction of $x$ to $A$ and $\neg A,$ respectively. Furthermore,  $(x_{-A},\alpha),$ denotes an act where $\alpha$ is placed on all coordinates in $A.$ See (Page 94, \cite{wa89}) for further details.} 
$A$ is \textit{essential} on $\mathbb R^{n}_{++}$ with respect to a weak order $\succsim$ if and only if $(x_{\neg A},v_{A})\succ (x_{\neg A},w_{A})$ for $x,v,w$ in $\mathbb R^{n}_{++}.$ A value function $V: \mathbb R^{n}_{++}\rightarrow \mathbb R$ is additive on $\mathbb R^{n}_{++}$ if there exist functions $\{v_{i}\}_{i=1}^{n},$ $v_{i}:\mathbb R\rightarrow \mathbb R,$ such that $V(x)=\sum\limits_{i=1}^{n}v_{i}(x_{i}).$ If $V$ is an additive representation for $\succsim$ then the functions $v_{i}$ are called additive value functions. For $\alpha,\beta, \gamma, \delta \in \mathbb R_{++}, $ $(\alpha,\beta)\succsim^{*}_{A}(\gamma,\delta)$ if there exist simple acts $x,y$ such that both $(x_{-A},\alpha)\succsim (y_{-A},\beta)$ and $(x_{-A},\gamma)\succsim (y_{-A},\delta).$

\prp{\nf(Wakker (1989))}
 \lb{wakker2}
 Let $n\geq 3,$ and let $\succsim$ be a binary relation on $\mathbb R^{n}_{++}.$ The following two statements are equivalent:
\ben[{\nf (a)}, topsep=3pt]
 \setlength{\itemsep}{-1pt} 
\ml There exists concave  nondecreasing nonconstant additive value functions $(v_{i})_{i=1}^{n}$ for $\succsim.$ 

\ml The binary relation $\succsim$ is a continuous weak order, weakly monotonic, every coordinate is essential, and furthermore, for all $i\in \{1,\ldots,n\}$, $(\alpha-\epsilon,\beta-\epsilon)\succsim_{i}^{*}(\alpha,\beta)$ whenever $(\alpha-\beta)\epsilon\geq 0.$ 
\een 
\prpp

Under weak order and weak monotonicity, we can apply Theorem \ref{thm_fequiv} to replace continuity of $\succsim$ by restricted solvability. In fact, any notion of continuity specified in Theorem \ref{thm_fequiv} can be applied.

\subsection{Chew-Karni's (1994) axiomatization of Choquet Expected Utility}

Chew-Karni's (1994) paper utilises a stronger version of the restricted solvability axiom  to obtain an axiomatization of subjective expected utility and Choquet expected utility models. In this subsection, we establish equivalences between their solvability axiom and the postulates appearing in Theorem 3.


Let $\Omega$ be a finite state space with $X\subseteq \Re$ denoting the set of consequences. An act $f$ is a function from $\Omega$ to $X$ with $F$ denoting the set of all acts. Given $f\in F, A\subset \Omega,$ and $c\in X,$ let $c_{A}f$ denote the act which yields $c$ on $A$ and  $f$ on $A^{c},$ the complement of $A.$ The authors use the following stronger version of the solvability axiom (their Axiom 6).

\df{\nf(Stronger RS)}
For all $f,g\in F,$ $A\subset \Omega,$ $c, c'\in X,$ if $c_{A}f\succsim g \succsim c'_{A}f,$ then there exists $c''\in X$ such that $g\sim c''_{A}f.$
\dff

\prp
Stronger RS implies restricted solvability and Wold continuity implies Stronger RS. Therefore, under conditions in Theorem 3, weak Wold continiuity is equivalent to  stronger RS.
\prpp

\rmk{\nf For the the arguments in the  first sentence, we do not need $X$ to be a subset of $\Re$. } 
\rmkk 

\subsection{Segal-Sobel's (2002) representation of min, max and sum}

\cite{ss02} provide three characterization theorems for preferences over $\mathbb R^{n}$ that can be represented by functionals of the form, $U(x_{1},\ldots,x_{n})$=$\min\{x_{i}\}_{i=1}^{n},$ $U(x_{1},\ldots,x_{n})$=$\max\{x_{i}\}_{i=1}^{n},$
$U(x_{1},\ldots,x_{n})$=$\sum\limits_{i=1}^{n} u(x_{i})$ where $u:\mathbb R\rightarrow \mathbb R,$ $U:\mathbb R^{n}\rightarrow \mathbb R.$ Under the assumption of weak monotonicity and order denseness, we can replace their continuity assumption by restricted solvability and restate their main theorem. Moreover, under continuity, our weak monotonicity implies \cite{ss02}'s monotonicity assumption. Below we provide their Theorem 3 as an application but our results apply to their other two theorems too, see \cite{ss02} for further details and background definitions.

\prp{\nf(\cite{ss02})} Let $n\geq 3$ and $\succsim$ satisfy completeness , order-denseness and transitivity. The following two conditions on $\succsim$ are equivalent.
\ben[{\nf (a)}, topsep=3pt]
 \setlength{\itemsep}{-1pt} 
\ml $\succsim$ satisfy restricted solvability, monotonicity, symmetry, comonotonic flatness, and partial separability.
\ml $\succsim$ can be represented by one of the following functions.
\ben[{\nf (a)}, topsep=1pt]
 \setlength{\itemsep}{-1pt} 
\ml $U(x_{1},\ldots,x_{n})=\min\{x_{1},\ldots,x_{n}\}$
\ml $U(x_{1},\ldots,x_{n})=\max\{x_{1},\ldots,x_{n}\}$
\ml $U(x_{1},\ldots,x_{n})=\sum\limits_{i=1}^{n} u(x_{i})$ for some strictly increasing u.
\een
\een
\prpp

\subsection{K\"obberling-Wakker's (2003) algebraic-topological distinction}

On the interplay between continuity and solvability in topological  and algebraic structures, \cite{kw03} note,

\bqu
We prefer the algebraic approach [solvability] to the topological approach [continuity] not only because its axioms are more general, but also because we consider these axioms to be more natural. 

\equ

\nt In our final application, using our Theorem 3, we demonstrate this interplay by restating the representation theorem of Cumulative Prospect Theory (CPT) using the solvability assumption, thus lending it empirical content without using the stronger assumption of continuity. For background definitions and details regarding CPT, readers may wish to see  \cite{kw03} for a comprehensive discussion. 
\prp
 Given that $X\subseteq R$ is a connected topological space and that $X^{n}$ is endowed with the product topology, assume $\succcurlyeq$ on $X^{n}$ be truly mixed. The following two statements are equivalent.
 
 \ben[{\nf (a)}, topsep=3pt]
 \setlength{\itemsep}{-1pt} 
 \ml CPT holds with a continuous utility function.
\ml $\succcurlyeq$ satisfies weak ordering, weak monotonicity, solvability, gain-loss consistency, sign-comonotonic tradeoff consistency.
\een

\prpp

\section{Concluding Remark}

 The theory and applications presented above, while of interest for its own sake, are not primarily motivated as an ivory-tower inquiry and aren't merely technical, simply to burnish the already strong foundations of partial equilibrium consumer theory and general  equilibrium Walrasian theory in mathematical economics on the one hand and issues related to measurement theory in mathematical psychology, on the other. The motivations which have led us to this inquiry are best summarized by the following extended quotation from \cite{ek12}.
 
 \bqu In recent years there has been a growing public fascination with the complex ``connectedness" of modern society. This connectedness is found in many incarnations: in the rapid growth of the Internet and the Web, in the ease with which global communication now takes place, and in the ability of news and information as well as epidemics and financial crises to spread around the world with surprising speed and intensity. These are phenomena that involve networks, incentives, and the aggregate behavior of groups of people; they are based on the links that connect us and the ways in which each of our decisions can have subtle consequences for the outcomes of everyone else. Drawing on ideas from economics, sociology, computing and information science, and applied mathematics, it describes the emerging field of study that is growing at the interface of all these areas, addressing fundamental questions about how the social, economic, and technological worlds are connected.
 \equ
 
 \nt It is the underscoring of this inter-disciplinarity and its urgency, and a corresponding bridging of the various communities, that constitutes the motivating engine of this work.

\bigskip

\appendix

\section*{Appendix}

In the Appendix, we present the proofs of the results and six technical examples.

\section{Proofs of the Results}\label{sec_proofs}

\prf[Proof of Theorem \ref{thm_basic}.]
	
{\bf (a)} The implication that Wold-continuity implies weak Wold-continuity follows from their definitions. \citet[Proposition 3]{uk19a} prove that  continuity implies Wold-continuity for finite dimensional spaces, their arguments extend to the current setting trivially.  



We next show that  weak Wold-continuity implies restricted solvability. Towards this end assume $\succsim$ is a weakly Wold-continuous binary relation on a convex set $X \subseteq \prod_{i\in I}X_{i}$.  
Let $y=(a_{i}, y_{-i} )$, $z= (b_{i},y_{-i} )$ and $y\succsim x\succsim z$.   
%
%
 If $y\sim x \sim z,$ or $y\succ x \sim z$, or $y\sim x \succ z,$ then restricted solvability trivially follows.  
Hence, assume  $y\succ x \succ z$. 
From weak Wold-continuity there exists a straight line $y\lambda z$ intersecting the indifference class of $x$, i.e., $\lambda y + (1-\lambda) z \sim x$,  
where $\lambda y + (1-\lambda) z= (\lambda a_{i}+(1-\lambda) b_{i}, y_{-i})$.
Therefore, $\succsim$ is  restricted solvable.

It remains to show that weak Wold-continuity implies Archimedean property. Towards this end assume $\succsim$ is Wold-continuity but not Archimedean. Then, there exists $x,y,z\in X$ such that $x\succ y$ but for all $\lambda \in (0,1)$, $x\lambda z\nsucc y$ (the proof of the case where $x\nsucc y\lambda z$ is analogous). By completeness, $y\succsim x\lambda z$. Pick $\lambda \in (0,1)$. If $y\succ x\lambda z$, then $x\succ y\succ  x\lambda z$ and weak Wold-continuity imply there exist $\delta \in (0,1)$ such that $y\sim x\delta z$.   If $y\sim  x\lambda z$, then set $\delta=\lambda$.  Then by transitivity, $x\succ x\delta z$. By weak Wold-continuity, there exists $\gamma\in (0,1)$ such that $x\succ x\gamma z\succ x\delta z\sim y$. Hence, by transitivity, $x\gamma z\succ y$. This furnishes us a contradiction with the assumption that for all $\lambda' \in (0,1)$, $x\lambda' z\nsucc y$. 

\medskip

\nt {\bf   (b)} Continuity implies separate continuity follows from the definitions of the two continuity postulates. For the proof of the remaining implications except  mixture-continuity implies separate continuity, see   \citet[Proposition 3]{uk19a}.


It remains to show that mixture-continuity implies separate continuity. Assume $\succsim$ is separately continuous. Pick an index $i$, $x\in X$ and a sequence $(y_i^n, y_{-i})$ converging $(y_i, y_{-i})\in X$ such that $(y_i^n, y_{-i})\succsim x$. mixture-continuity implies that the set $A=\{\lambda\in [0,1]~|~(y_i, y_{-i})\lambda y\succsim x\}$ is closed. It is clear that $1\in A$ and for all $\lambda >0$, $\lambda \in A$. Hence $1\in A$. Therefore, The proof that for all $x\in X$ and all lines $L$ parallel to any coordinate axis,  $A_\succsim(x)\cap L$ is closed. 
%
The proof that for all $x\in X$ and all lines $L$ parallel to any coordinate axis,  $A_\precsim(x)\cap L$ is closed analogously follows. 
Since $\succsim$ is complete, $\succsim$ is separately continuous.


\medskip

\nt {\bf  (c)}   Assume $\succsim$ is separately continuous.  Let $ (a_{i}, y_{-i})\succsim x\succsim (b_{i}, y_{-i})$. Separate continuity implies that  the weakly worse-than and weakly better-than sets of $x$ are closed in the subspace of $X$ determined by the line $L_i$ parallel to the i-$\text{th}$ coordinate axis and passing through $y_{-i}$, i.e. $\succsim_{x}\upharpoonright_{L_i}$ and $\precsim_{x}\upharpoonright_{L_i}$ are closed in $X\cap L_i$. Since $X\cap L_i$ is convex, it is connected. Moreover, it follows from the completeness of the relation that the intersection of these two sets is non-empty, i.e., $ \succsim_{x}\upharpoonright_{L_i}\cap \precsim_{x}\upharpoonright_{L_i}\neq \varnothing$. Hence, there exists $\tilde x\in X\cap L_i$ such that $\tilde x\sim x$. Since $\tilde x\in X\cap L_i$, there exists $c_i\in \Re$ such that $\tilde x=(c_i,\bar y_{-i})$. Therefore, $\succsim$ is restricted solvable.

\medskip

\nt {\bf  (d)} Assume $\succsim$ is unrestricted solvable. And assume its is not restricted solvable.  Then  there exists no $b$ such that $bq\sim ap$ whenever $\bar{b}q\succsim ap \succsim \uline{b}q$. But this contradicts unrestricted solvability.  
\prff


\prf[Proof of Theorem \ref{thm_bequiv} ]
We provide the proof by induction. Note that for $n=1$, separate continuity is equivalent to continuity by definition. Now let $n=2$ and assume without loss of generality that $\succsim$ is weakly monotone in its first coordinate.  We now show that $\succsim$ has closed lower sections. Towards this end, assume there exist $x\in X$ and a sequence $y^k\ra y$ such that $y^k\precsim x$ for all $k$ and $y\nprecsim x$. Then, either $y\succ x$ or $y\bowtie x$.   
 
Assume $y\succ x$.  
 Separate continuity implies that there exists $\varepsilon>0$ such that for all $z$ in the $\varepsilon$-neighborhood of $y$, $(z_1, y_2)\succ x$. 
 Pick $(z_1, y_2)$ in the $\varepsilon$-neighborhood of $y$ such that $z_1<y_1-\varepsilon/2$. Separate continuity then implies that  there exists  $\delta_1>0$ such that for all $z'$ in the $\delta_1$-neighborhood of $(z_1, y_2)$, $(z_1, z'_2)\succ x$.   
Define $\delta=\min \{\varepsilon/2, \delta_1\}$. Since $y^k\ra y$, there exists $\hat k\in \mathbb N$ such that $y^{\hat k}$ is in the $\delta$-neighborhood of $y$.   Then, $(z_1, y_2^{\hat k})$ is in the $\delta_1$-neighborhood of $(z_1, y_2)$ and  $z_1<y_1^{\hat k}$.  Hence, by weak monotonicity, $y^{\hat k}\succsim (z_1, y_2^{\hat k})$. It follows from $(z_1, y_2^{\hat k})\succ x$ and transitivity of $\succsim$ that\fn{Note   $\succ$ is transitive; see \cite{ku21et} for details on the relationship between  different transitivity postulates.} $y^{\hat k}\succ x$. This furnishes us a contradiction.

 When $y\bowtie x$, replacing $\succ$ with $\bowtie$ in the   paragraph  above  yields $y^{\hat k}\succsim (z_1, y_2^{\hat k})$, as above. Then,  $y^{\hat k}\succsim (z_1, y_2^{\hat k})\bowtie x$. If $x\succsim y^{\hat k}$, then transitivity of $\succsim$ implies that $x  \succsim (z_1, y_2^{\hat k})$, yielding a contradiction. Hence $x\nsuccsim y^{\hat k}$. This furnishes us a contradiction. 
 
 Therefore, $\succsim$ has closed lower sections.  The proof of the closed upper sections is analogous. Moreover, the proof of the open sections have a similar construction.  Finally, completeness of $\succsim$ follows from \citet[Theorem 2]{ku21et}.

Now assume the theorem is true for  $n-1$ dimensional spaces, $n>2$. We next show that it is true for   $n$ dimensional spaces. Recall that $\succsim$ is weakly monotone in all   coordinates (except possibly one).  Pick a coordinate $i$ in which $\succsim$ is weakly monotone. We now show that $\succsim$ has closed lower sections. Towards this end, assume there exist $x\in X$ and a sequence $y^k\ra y$ such that $y^k\precsim x$ for all $k$ and $y\nprecsim x$. Then, either $y\succ x$ or $y\bowtie x$.   
 
Assume $y\succ x$.  
 Separate continuity implies that there exists $\varepsilon>0$ such that for all $z$ in the $\varepsilon$-neighborhood of $y$, $(z_i, y_{-i})\succ x$. 
 Pick $(z_i, y_{-i})$ in the $\varepsilon$-neighborhood of $y$ such that $z_i<y_i-\varepsilon/2$. By induction hypothesis, $\succsim$ is continuous on $n-1$ dimensional spaces. Hence,  there exists  $\delta_i>0$ such that for all $z'$ in the $\delta_i$-neighborhood of $(z_i, y_{-i})$, $(z_i, z'_{-i})\succ x$.   
Define $\delta=\min \{\varepsilon/2, \delta_i\}$. Since $y^k\ra y$, there exists $\hat k\in \mathbb N$ such that $y^{\hat k}$ is in the $\delta$-neighborhood of $y$.   Then, $(z_i, y_{-i}^{\hat k})$ is in the $\delta_i$-neighborhood of $(z_i, y_{-i})$ and  $z_i<y_i^{\hat k}$.  Hence, by weak monotonicity, $y^{\hat k}\succsim (z_i, y_{-i}^{\hat k})$. It follows from $(z_i, y_{-i}^{\hat k})\succ x$ and transitivity of $\succsim$ that  $y^{\hat k}\succ x$. This furnishes us a contradiction.

The remaining part of the proof is analogous to the case $n=2$ and uses the modification of the construction above for general $n$, hence omitted. 
\prff

\prf[Proof of Theorem \ref{thm_fequiv}]
 Let $\succsim$ be a complete, transitive, order dense and weakly monotonic binary relation on a convex and order-bounded set $X \subseteq \mathbb R^n$. Theorems  \ref{thm_basic} and \ref{thm_bequiv}  prove all  the relationships except that restricted solvability implies separate continuity. Towards this end, assume $\succsim$  is  restricted solvable but not separately continuous. 
 
 
 Assume there exists $x\in X$ and a line $L$ parallel to a coordinate axis such that $A_\succsim(x)\cap L$ is not closed. 
Then, there exist $x\in X$, an index $i$, and a sequence $y^k \ra y$ on the line $L_i$ parallel to coordinate $i$ such that $y^k\succsim x$ for all $k$ and $y\prec x$. Then it follows from transitivity and weak monotonicity assumptions that $y_i^k> y_i$ for all $k$. Pick $m\in \mathbb N$.   
Since $y\prec x$, order denseness implies that there exists $x'\in X$ such that $y\prec x'\prec x$.  Then  transitivity implies $y\prec x'\prec y^m$, and hence restricted solvability implies there exists $z\in L_i$ such that $z\sim x'$. It follows from weak monotonicity that $y_i<z_i<y^m_i$, and from transitivity that for all $z_i'\in (y_i, z_i]$, $(z'_i, z_{-i})\prec x$.  However, since $y^k\ra y$, there exists $z_i'\in (y_i, z_i]$, such that $z_i'=y_i^k$ for some $k$ and $(z'_i, z_{-i})\succsim x$. This yields a contradiction.  
 
 The proof that for all $x\in X$ and all lines $L$ parallel to any coordinate axis,  $A_\precsim(x)\cap L$ is closed analogously follows. 
 Since $\succsim$ is complete, $\succsim$ is separately continuous. 
\prff


\begin{proof}[Proof of Theorem \ref{thm_fequivInf}] 
We first show that Archimedean postulate implies mixture-continuity. Assume there exist $x,y,z\in X$ such that $\{\lambda: x\lambda y\succsim z\}$ is not closed.    Then there exists $\lambda^n\ra \lambda$ such that $x\lambda^n y\succsim z$ but $z\succ x\lambda y$. By strong order-boundedness assumption, there exists $\varepsilon>0$ such that $p=((x\lambda y)_i +\varepsilon)_{i\in I}\in X$. By weak monotonicity, $p\succsim p\delta (x\lambda y)$ for all $\delta\in [0,1]$. By Archimedean, there exists $\beta\in (0,1)$ such that $z\succ   p\beta (x\lambda y)$. Note that there exists $\epsilon \in (0,\varepsilon)$ such that $(p\beta (x\lambda y))_i - (x\lambda y)_i > \epsilon$ for all $i\in I$. Therefore, there exists $N\in \mathbb N$ such that for all $n\geq N$, $(x\lambda^n y)_i < (p\beta (x\lambda y))_i$ for all $i\in I$. By weak monotonicity, $p\beta (x\lambda y)\succsim x\lambda^n y$ for all $n\geq N$. By transitivity, $z\succ x\lambda^n y$ for all $n\geq N$. This yields a contradiction. Closedness of the lower sections follows analogously.

By the figure above, we now prove the equivalence of mixture continuity, Archimedean and weak Wold-continuity.  The proof that restricted solvability implies separate continuity follows from the argument in the proof of  Theorem 4 since it does not hinge on the dimension of the underlying space.  
 \end{proof}

\section{Examples}
\label{sec_examples}

In this section we show that in Figure 1, the converse relationships among the continuity postulates are false.  All the binary relations presented here satisfy completeness and transitivity.

The first example shows that continuity, and hence all other continuity postulates, do not imply unrestricted solvability even under  weakly monotone and convex preferences.    

\vspace{0.3cm}

\nt \textbf{1.} Let  $\succsim$ be a   binary relation defined on $\mathbb R^{2}$ as
 $(x_{1},x_{2})\succsim (y_{1},y_{2})$ if and only if  $f(x_{1},x_{2})\geq f(y_{1},y_{2})$ where $f(x_{1},x_{2})=x_{2}.$  
%
In order to see that $\succsim$ does not satisfy unrestricted solvability, for all $x_1\in \Re$, $(2,2)\succ (x_1,1)$, hence $\succsim$ is not unrestricted solvable in the second component. It is clear that $\succsim$ is continuous, hence it satisfies the remaining six continuity postulates, following from Theorem \ref{thm_basic}.
\qed

\medskip

The second example shows that unrestricted solvability and restricted solvability do not imply separate continuity, Archimedean and weak Wold-continuity, and hence any of the other continuity postulates. 

\smallskip

\nt \textbf{2.} Let the binary relation $\succsim$ be defined on a bounded subset $X\subset\mathbb R^{2}$ as $(x_{1},x_{2})\succsim (y_{1},y_{2})$ if and only if $f(x_{1},x_{2})\geq f(y_{1},y_{2})$  
where 
\[
f(x_{1},x_{2}) = 
\begin{cases}
0, & \text{ if } x_{1}+x_{2}< 1 \\
0.8, & \text{ if }x_{1}+x_{2}= 1 \\
1, & \text{ if } x_{1}+x_{2}> 1
\end{cases}
\]


To see why this is not separately continuous and hence not continuous, pick $(x_{1},x_{2})=(1,0).$ For the binary relation to be separately continuous, the restriction of $(1,0)$ to any line parallel to either $x_{1}$ or $x_{2}$ axis should be continuous. Let $x_{2}=0.5.$ Then $\succsim_{(1,0)}\upharpoonright_{x_{2}=0.5}=(0.5,b]$ where $b$ is some bound on $x_{1}.$ 
This representation is neither Wold nor weak Wold-continuous (does not satisfy order-denseness). However, this binary relation is both restricted and unrestricted solvable. 

This is not Archimedean. Pick $x=(0.6,0.4)$ and $y=z=(0.6,0.5).$ Then $y\succ x$ but $y\sim x\delta z$ for any $\delta\in (0,1).$
\qed

\vspace{0.3cm}

The third example shows that separate continuity does not imply weak Wold-continuity, Archimedean and unrestricted solvability, hence does not imply mixture-continuity, Wold-continuity and continuity postulates. It also shows that restricted solvability does not imply any of these continuity postulates. 
\medskip

\nt \textbf{3.} Let the binary relation $\succsim$ be defined on $\mathbb R^{2}$ as $(x_{1},x_{2})\succsim (y_{1},y_{2})$ if and only if $f(x_{1},x_{2})\geq f(y_{1},y_{2})$ where 
\[
f(x_{1},x_{2}) = 
\begin{cases}
\dfrac{x_{1}x_{2}}{x_{1}^{2}+x_{2}^{2}}  & \text{ if } (x_{1},x_{2})\neq (0,0),  \\
  0   & \text{ if } (x_{1},x_{2}) = (0,0).  \\
\end{cases}
\]



In this example, we have, $(1,1)\succ (3,1)\succ (0,0).$ There exists no $\lambda \in [0,1]$ such that $\lambda (1,1)+(1-\lambda)(0,0) \sim (3,1).$ Hence, this is not Weak Wold-continuous. Consequently, this is also not Wold-continuous.  

It is easy to show that this binary relation is not unrestricted solvable. Pick a pair $(1,1)$ and let the third element of the to-be-determined pair is $0.$ Then unrestricted solvability claims that there exists $q$ such that $(1,1)\sim (q,0).$ In this case, there exists no $q$ such that the indifference holds.

Finally this example illustrates that a binary relation that is restricted solvable, but not Archimedean.   The function is continuous on any line parallel to a coordinate axis, hence the induced binary relation is restricted solvable. However, it is not continuous on the 45-degree line. In particular it is defined on $\Re^2_+$ and the function's value is 1 on all point on the 45-degree line except 0, and at 0 its value is 0. Therefore if you pick $x=(0,0), y=z=(1,1)$, then $y\succ x$ but for all $\lambda\in (0,1)$, $y\sim x\lambda z$, hence Archimedean property fails. Similarly, it also shows that Archimedean does not imply mixture-continuity. 
   \qed

The fourth example illustrates a binary relation that is Archimedean  and Wold-continuous, and hence weakly Wold-continuous  and restricted solvable, but not mixture-continuous, separately continuous and continuous. It is originally presented in \cite{uk19a}.  

\vspace{0.3cm}

\nt \textbf{4.}
 Let $X=\Re_+$ and $f(x)=\text{sin}(1/x)$ if $x>0$ and $f(0)=1$. Then, it is clear that $f(x)\in [-1,1]$ for all $x\in X$. Define a binary relation $\succsim$ on $X$ as  $x\succsim y$ is and only if $f(x)\geq f(y)$. Pick $\bar x$ such that $f(\bar x)\in (0,1)$. The set $\{x'\in X|\bar x\succsim x'\}$ is not closed since it contains a sequence $x_n\ra 0$ but $0\succ \bar x$. Therefore, $\succsim$ is not continuous. Moreover, since $X$ is one dimensional, $\succsim$ is not mixture-continuous and separately continuous. See \citet[Proof of Proposition 3]{uk19a} for a proof that $\succsim$ satisfies Wold-continuity and Archimedean properties. 
%
   \qed

\medskip

The fifth example shows that Archimedean property does not imply restricted solvability, hence all other continuity postulates in Figure 1.
\smallskip

\nt {\bf 5.} Let  $\succsim$ be a  binary relation  defined on $\mathbb R^{2}_+$ as $(x_{1},x_{2})\succsim (y_{1},y_{2})$ if and only if $f(x_{1},x_{2})\geq f(y_{1},y_{2})$ where 
\[
f(x_{1},x_{2}) = 
\begin{cases}
 \text{sin}(1/x_2)  & \text{ if } x_2>0,\\
 x_1  & \text{ if } x_1\in[0,1]\cap \mathbb Q, x_2=0,\\
 0  & \text{ if } x_1\in[0,1]\cap \mathbb Q^c, x_2=0,\\
 1  & \text{ if } x_1> 1, x_2=0.\\
\end{cases}
\]

It is not difficult to show that $\succsim$ satisfies the Archimedean property.  
%
In order to see that $\succsim$ fails to satisfy restricted solvability, note that $(1,0)\succ (0,2) \succ (0,0)$ since $f(1,0)=1, f(0,0)=0$ and $f(0,2)$ is an irrational number in (0,1) interval (sin() of a non-zero rational number is irrational). Hence, for all $x_1\in \Re_{+}$,  either $(x_1,0)\succ (0,2)$ or $(x_1,0)\prec (0,2)$. 
\qed

\medskip

The sixth and final example illustrates the importance of the interiority assumption in our results by showing that Theorems 2 and 3 above are false on a general choice set; see also \cite{ks15} for the use of the interiority assumption to show that the Archimedean property is equivalent to mixture-continuity under cone-monotonicity.
\smallskip

\nt {\bf 6.} 
Let $X=\{x\in (-1,1)^2: x_1=x_2\}$. Then $X$ is an order bounded and convex set in $\Re^2$, it is not open in $\Re^2$. (Note that $X$ is open in the one dimensional subspace containing it.) Let $u: X\rightarrow \Re$ be defined by $u(x)=0$ for all $x\leq 0$ and $u(x)=1$ if $x_1>0$. Let $\succsim$ be the preference relation induced by $u$. Then, $A_\succsim (0.5, 0.5)=\{x\in X: x_1>0\}$ which is not closed in $X$, hence $\succsim$ is not continuous. It is clear that $\succsim$ is weakly monotone. Since the restriction of any line parallel to a coordinate axis is a singleton,  $\succsim$ is trivially separately continuous.  
 
 Note that in the example above, we can replace $X$ by $\{x\in [-1,1]^2: x_1=x_2\}$, hence boundaries can be included. Similarly, we can replace $X$ by a set with non-empty interior: $\{x\in [0,1]^2: 2x_1\geq x_2\geq 0.5 x_1 \}$.  Let $u: X\rightarrow \Re$ be defined by $u(0)=0$ and $u(x)=1$ if $x \neq 0$. Let $\succsim$ be the preference relation induced by $u$. Then $\succsim$ is weakly monotone, separately continuous and discontinuous.

\bigskip

\setlength{\bibsep}{5pt}
\setstretch{1.05}

%
%
%
%
%
%

\ifx\undefined\BySame
\newcommand{\BySame}{\leavevmode\rule[.5ex]{3em}{.5pt}\ }
\fi
\ifx\undefined\textsc
\newcommand{\textsc}[1]{{\sc #1}}
\newcommand{\emph}[1]{{\em #1\/}}
\let\tmpsmall\small
\renewcommand{\small}{\tmpsmall\sc}
\fi

\end{document}